\documentclass[
  ,final            
  ,numberedheadings 
  ]
  {aipproc}

\layoutstyle{6x9}

\begin{document}

\title[Energy and rapidity dependence of electric charge correlations (NA49)]{Energy and rapidity dependence of electric charge correlations at 20-158GeV beam energies at the CERN SPS (NA49)}

\classification{<>}
\keywords      {Balance Function,correlations}

\author{P. Christakoglou, A. Petridis, M. Vassiliou for the NA49 collaboration}{
  address={University of Athens}
}

\begin{abstract}

Electric charge correlations are studied with the Balance Function method for central Pb + Pb collisions at the CERN - SPS. The results on centrality selected Pb + Pb interactions at 40 and 158 AGeV are presented for the first time for two different rapidity intervals. In the mid-rapidity region a decrease of the width with increasing centrality of the collision is observed whereas in the forward rapidity region this effect vanishes. This could suggest a delayed hadronization scenario. In addition, the results from a first attempt to study the energy dependence of the Balance Function throughout the whole SPS energy range, are presented. The suitably scaled decrease of the width is approximately constant for the intermediate energies (30 to 80 AGeV) and gets stronger for the highest SPS and RHIC energies. On the other hand, both URQMD and HSD simulation results show no dependence on the collision energy.

\end{abstract}

\maketitle

\section{Introduction}

The study of correlations and fluctuations is expected to provide additional information on the reaction mechanism of high energy nuclear collisions \cite{QM,Alb94,QGP,QM04}. In particular, many proposed event-by-event fluctuation observables have already been analyzed \cite{Stock99}. More recently, another important measure of correlations, the Balance Function (BF), was introduced by Bass, Danielewicz and Pratt \cite{Pratt}. It measures the correlation of the oppositely charged particles produced during a heavy ion collision and its width can be related to the time of hadronization. The BF is derived from the charge correlation function that was used to study the hadronization of jets in p+p collisions at the ISR \cite{Drij} and $e^-+e^+$ annihilations at PETRA \cite{Aih}. The first results on the BF were obtained for Au+Au collisions by the STAR collaboration at RHIC \cite{STAR} and for Pb+Pb collisions by the NA49 collaboration at the CERN-SPS \cite{BF_NA49}.

The motivation for studying the Balance Function comes from the idea that hadrons are produced locally as oppositely charged particle pairs. Particles of such a pair are separated in rapidity due to the initial momentum difference and secondary interactions with other particles. Particles of a pair that were created earlier are separated further in rapidity because of the expected large initial momentum difference and the long lasting rescattering phase. On the other hand, oppositely charged particle pairs that were created later are correlated within a smaller interval $\Delta y$ of the relative rapidity. Our aim is to measure the degree of this separation of the balancing charges and to find possible indications for delayed hadronization.

In order to examine the pseudo-rapidity ($\eta$) correlation of charged particles the BF is defined as the difference of the correlation function of oppositely charged particles and the correlation function of like-charge particles normalized to the total number of particles. The definition of the BF reads \cite{Pratt}:

\begin{center}
\begin{equation}
B(\Delta \eta) = \frac{1}{2} \Big[ \frac{N_{+-}(\Delta \eta) -
N_{--}(\Delta \eta)}{N_{-}} + \frac{N_{-+}(\Delta \eta) -
N_{++}(\Delta \eta)}{N_{+}}  \Big].
\label{BF_DEF2}
\end{equation}
\end{center}

\vspace{0.2 cm}

The most interesting property of the BF is its width. Early stage hadronization is expected to result in a broad BF, while late stage hadronization leads to a narrower distribution \cite{Pratt}. The width of the BF can be characterized by the weighted average $\langle \Delta \eta \rangle$:

\begin{center}
\begin{equation}
\langle \Delta \eta \rangle = \sum_{i=0}^k{(B_i \cdot \Delta \eta _i)}/\sum_{i=0}^k{B_i},
\label{width}
\end{equation}
\end{center}

\noindent where \emph{i} is the bin number of the BF histogram.

In the following sections we will first describe in brief the NA49 experimental setup. The next two sections will be dedicated to the new NA49 experimental results on the rapidity and the energy dependence of the Balance Function. We will conclude with the summary.

\section{Experimental Setup}

\begin{figure}
  \includegraphics[height=.3\textheight]{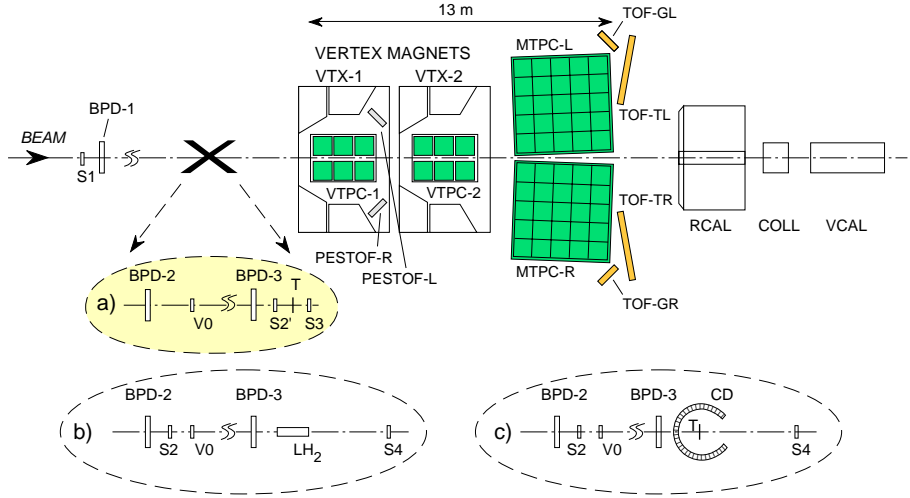}
  \caption{The NA49 experimental setup.}
  \label{na49_setup}
\end{figure}

The NA49 detector \cite{na49_nim} is a wide acceptance hadron spectrometer for the study of hadron production in collisions of hadrons or heavy ions at the CERN SPS. The main components are four large - volume Time Projection Chambers (TPCs) (Fig. \ref{na49_setup}) which are capable of detecting 80\% of some 1500 charged particles created in a central Pb+Pb collision at 158\emph{A} GeV. 

The targets are C (561 mg/cm$^{2}$), Si (1170 mg/cm$^{2}$) disks and a Pb (224 mg/cm$^{2}$) foil for ion collisions and a liquid hydrogen cylinder (length 20 cm) for hadron interactions. They are positioned about 80 cm upstream from VTPC-1.

The centrality of a collision is selected (on-line for central Pb+Pb, Si+Si and C+C and off-line for minimum bias Pb+Pb, Si+Si and C+C interactions) by a trigger using information from a downstream calorimeter (VCAL), which measures the energy $E_0$ of the projectile spectator nucleons.

\section{Rapidity Dependence}

In order to investigate the properties of hadronization in heavy ion collisions, as proposed by the BF methodology \cite{Pratt}, we studied the system size and centrality dependence of the width of the BF at two SPS energies ($\sqrt{s} = 17.2$ GeV and $\sqrt{s} = 8.8$ GeV) and in two different pseudo-rapidity intervals. The defined pseudo-rapidity regions for the highest energy are $2.6 \leq \eta \leq 3.9$ named as mid-rapidity region and $4.0 \leq \eta \leq 5.4$ as forward rapidity region. The corresponding pseudo-rapidity regions for the second energy are $1.8 \leq \eta \leq 3.2$ (mid-rapidity region) and $3.3 \leq \eta \leq 4.7$ (forward rapidity region).

\begin{figure}
\includegraphics[height=.3\textheight]{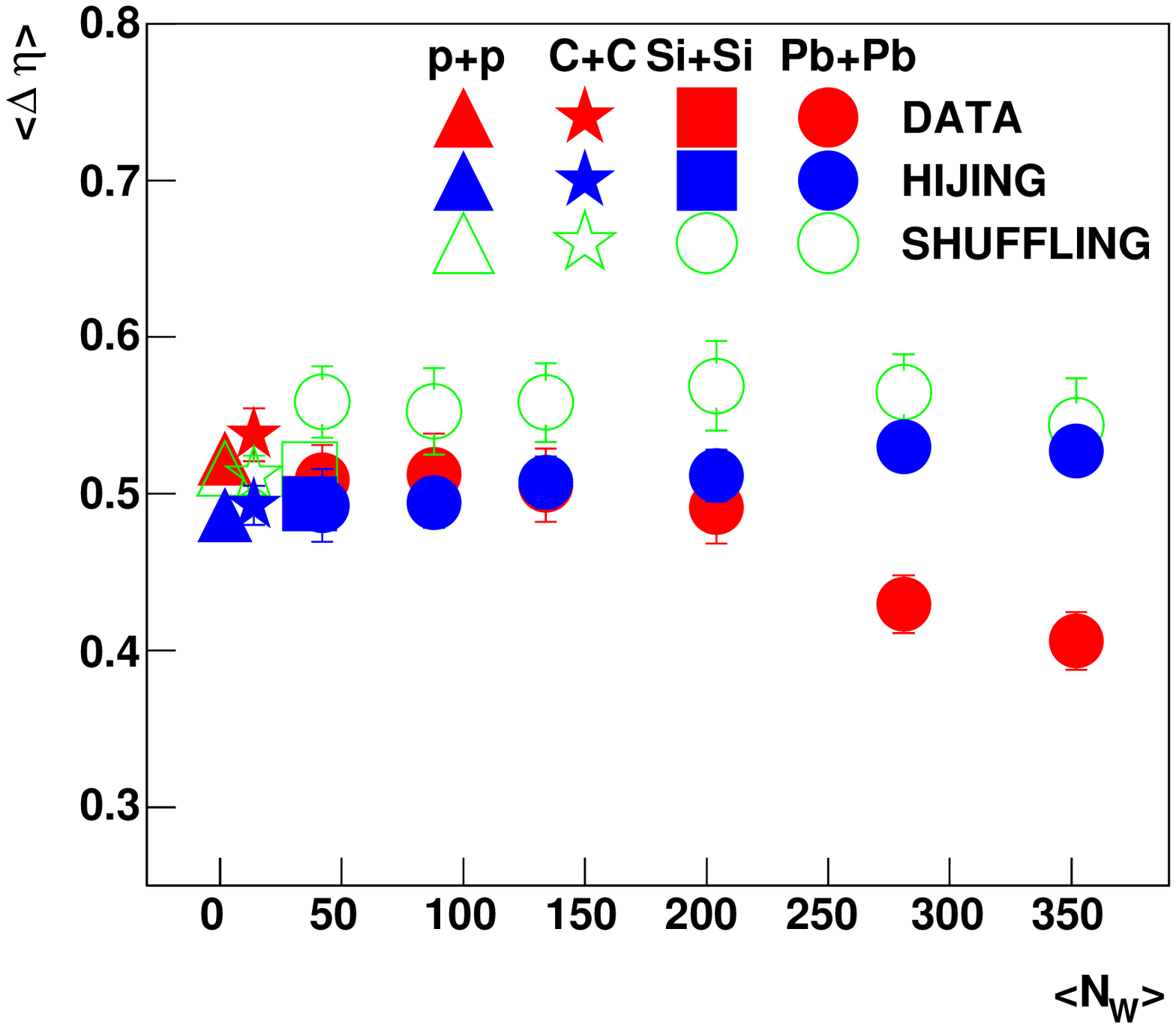}
\includegraphics[height=.3\textheight]{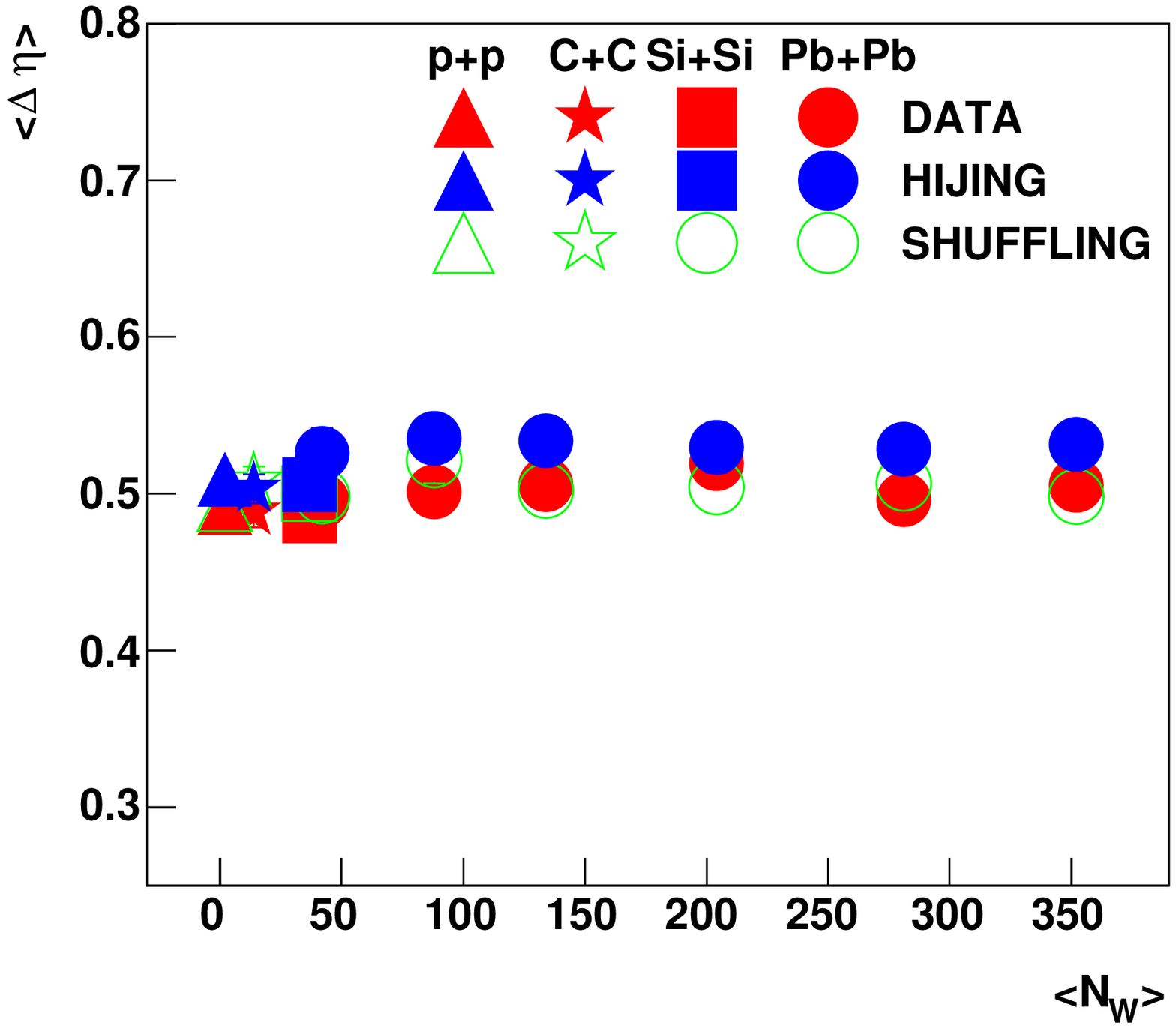}
\caption{The system size and centrality dependence of the measured width of the Balance Function for charged particles at $\sqrt{s_{NN}} = 17.2$ GeV as a function of the mean number of wounded nucleons for the two different rapidity intervals analyzed: the mid-rapidity region ($2.6 \leq \eta \leq 3.9$ - left plot) and the forward rapidity region ($4.0 \leq \eta \leq 5.4$ - right plot). }
\label{NA49_rap_160}
\end{figure}

\begin{figure}
\includegraphics[height=.3\textheight]{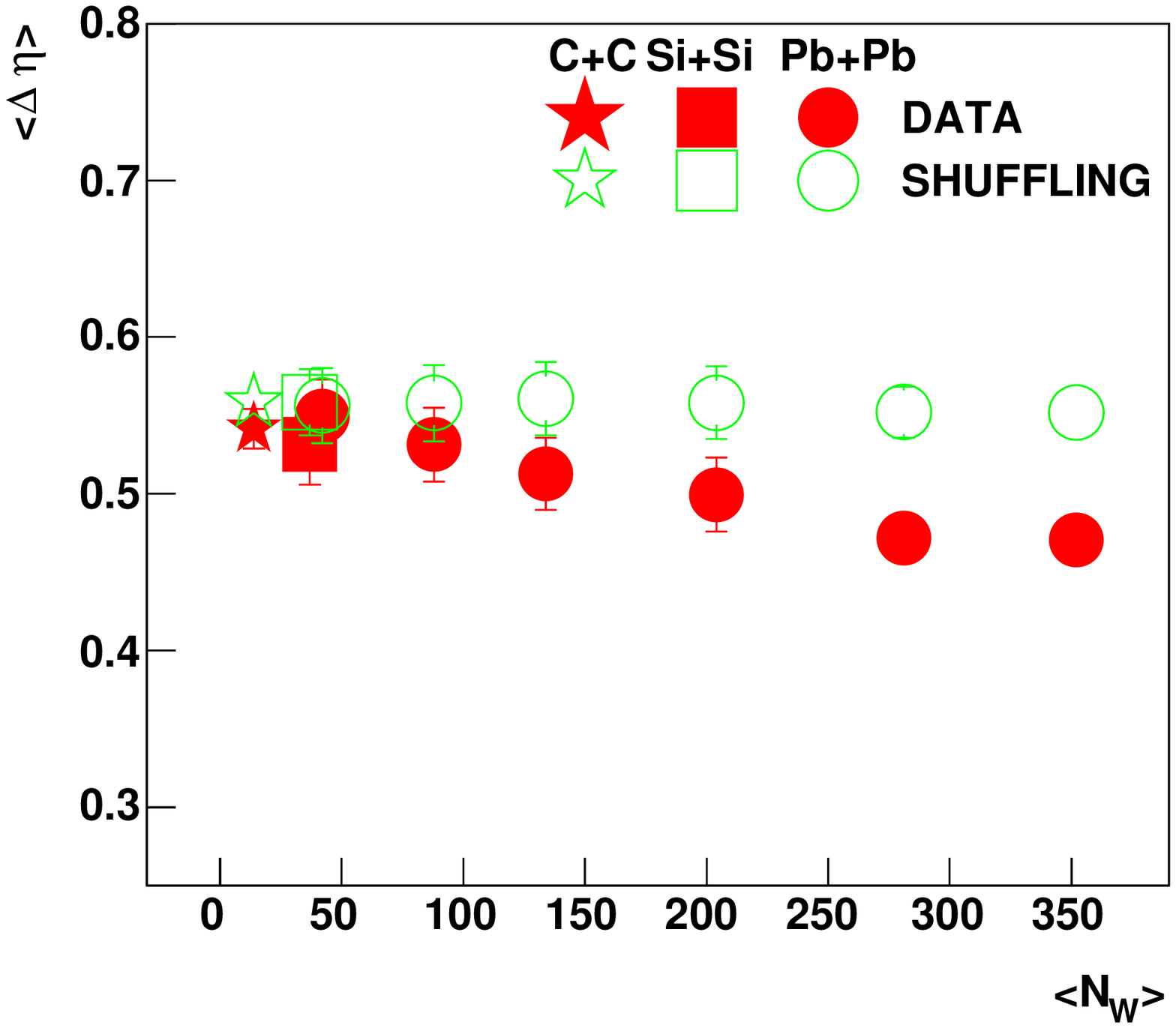}
\includegraphics[height=.3\textheight]{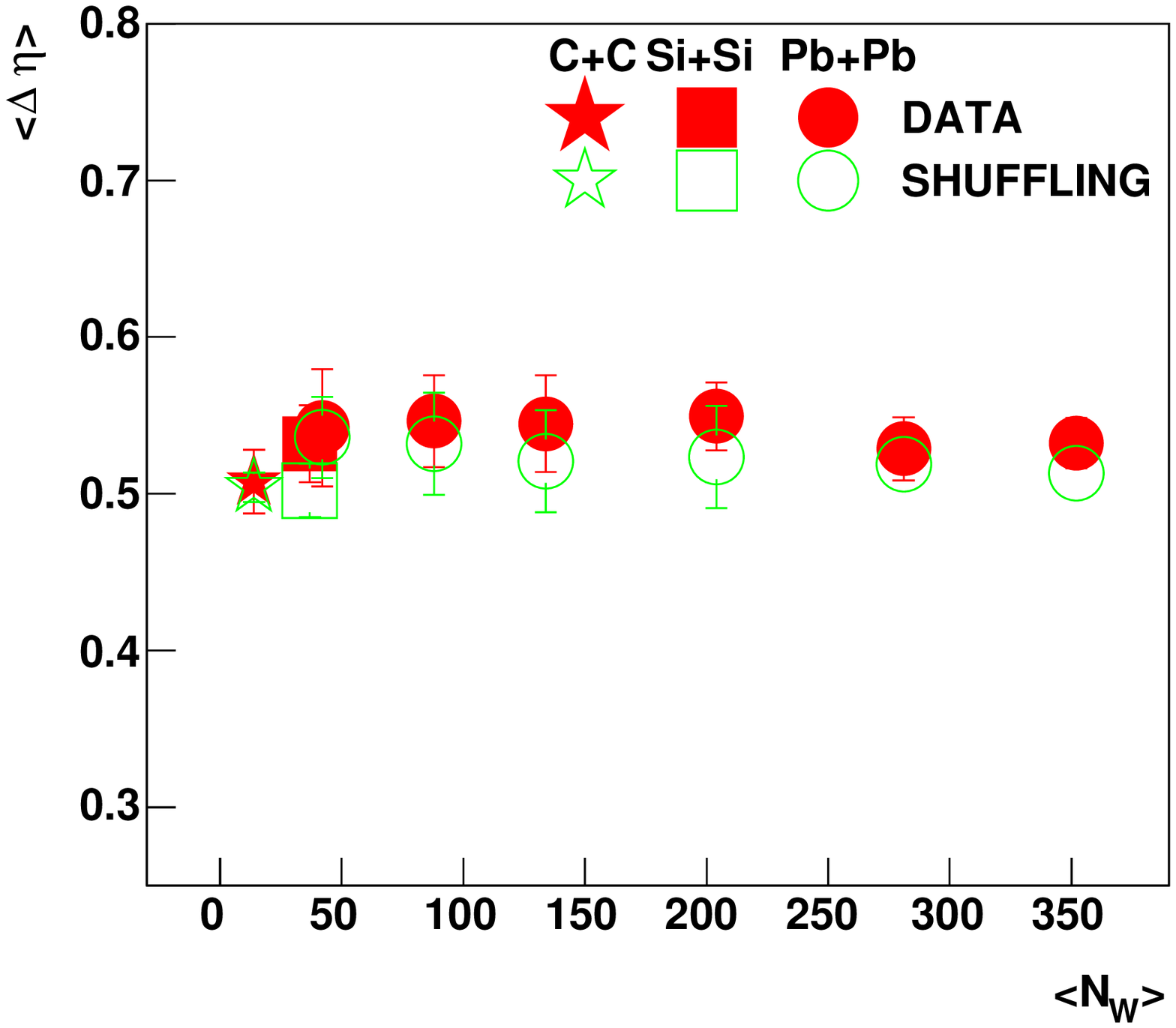}
\caption{The system size and centrality dependence of the measured width of the Balance Function for charged particles at $\sqrt{s_{NN}} = 8.8$ GeV as a function of the mean number of wounded nucleons for the two different rapidity intervals analyzed: the mid-rapidity region ($1.8 \leq \eta \leq 3.2$ - left plot) and the forward rapidity region ($3.3 \leq \eta \leq 4.7$ - right plot). }
\label{NA49_rap_40}
\end{figure}

Fig. \ref{NA49_rap_160} shows the width of the BF distributions for real, HIJING \cite{Hijing} and shuffled data as a function of the mean number of wounded nucleons, for the two pseudo-rapidity regions analyzed (mid-rapidity region-left plot, forward rapidity region-right plot) for $\sqrt{s_{NN}} = 17.2$ GeV. There is a clear decrease of the width with increasing centrality for real data in the mid-rapidity region while this dependence vanishes in the forward region. The same behavior can be observed for $\sqrt{s_{NN}} = 8.8$ GeV in fig. \ref{NA49_rap_40}.

For the smaller systems, such as p+p, C+C, Si+Si, there is no significant difference in the widths between the two rapidity regions at both energies. A difference in the widths appears only in the more central Pb+Pb collisions for both cases.

\section{Energy Dependence}

Finally, an attempt was made to study the energy dependence of the Balance Function with the NA49 detector in the most central Pb+Pb events over the whole available SPS energy range. These data samples were also passed through the shuffling mechanism \cite{STAR,BF_NA49} to obtain an estimate of the highest possible value of the width for each energy. The pseudo-rapidity interval analyzed for each energy was limited to the same range (1.4 units) and was located around mid-rapidity. In order to quantify the decrease of the width for the different energies, we introduced the normalized parameter W which is defined by the following equation:

\begin{center}
\begin{equation}
W = \frac{100 \cdot (\langle \Delta \eta \rangle_{shuffled} - \langle \Delta \eta \rangle_{data})}{\langle \Delta \eta \rangle_{shuffled}}.
\label{W}
\end{equation}
\end{center}

In other words this parameter indicates how many percent the measured width differs from the estimated shuffled value.

\begin{figure}
\includegraphics[height=.3\textheight]{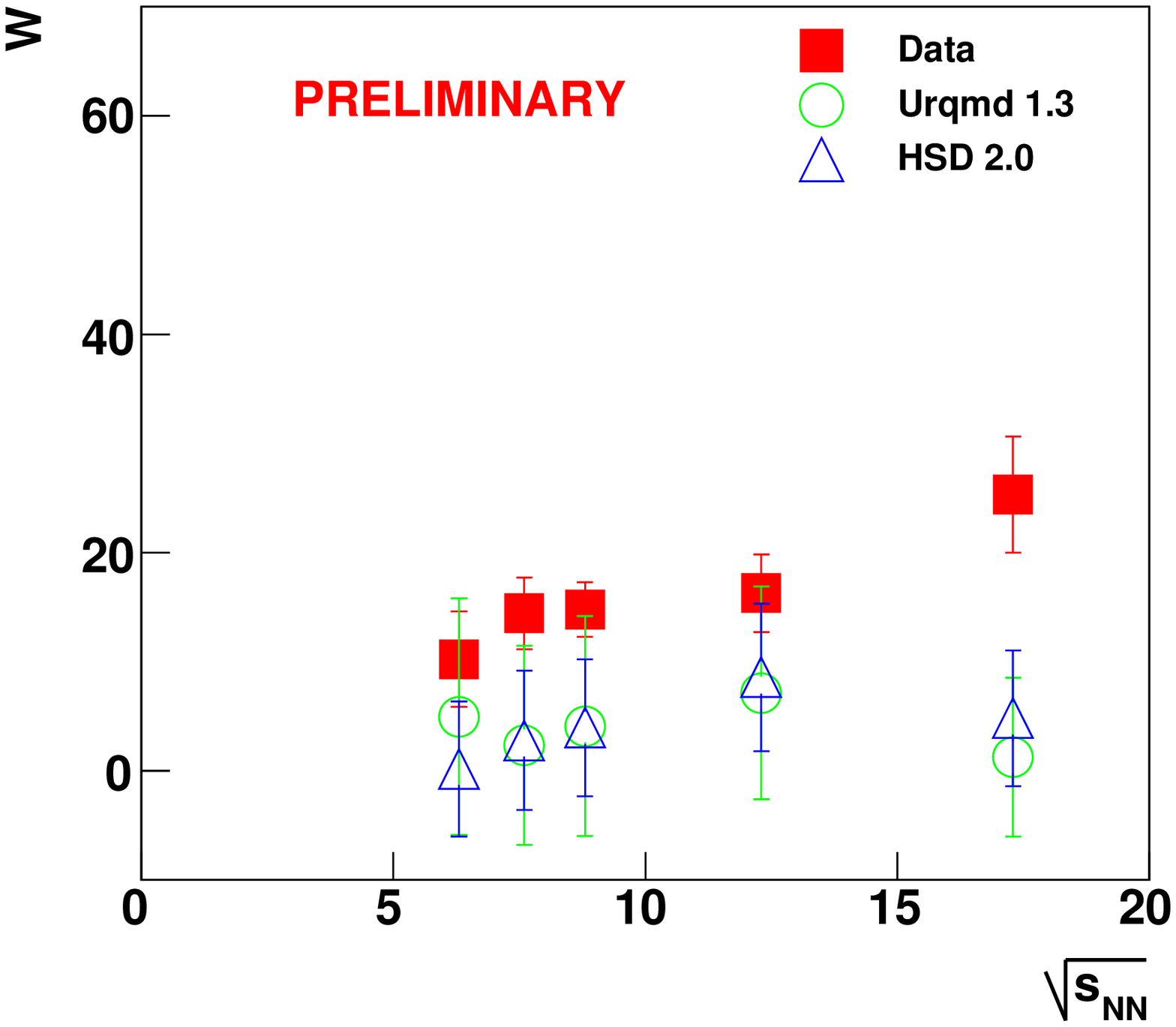}
\includegraphics[height=.3\textheight]{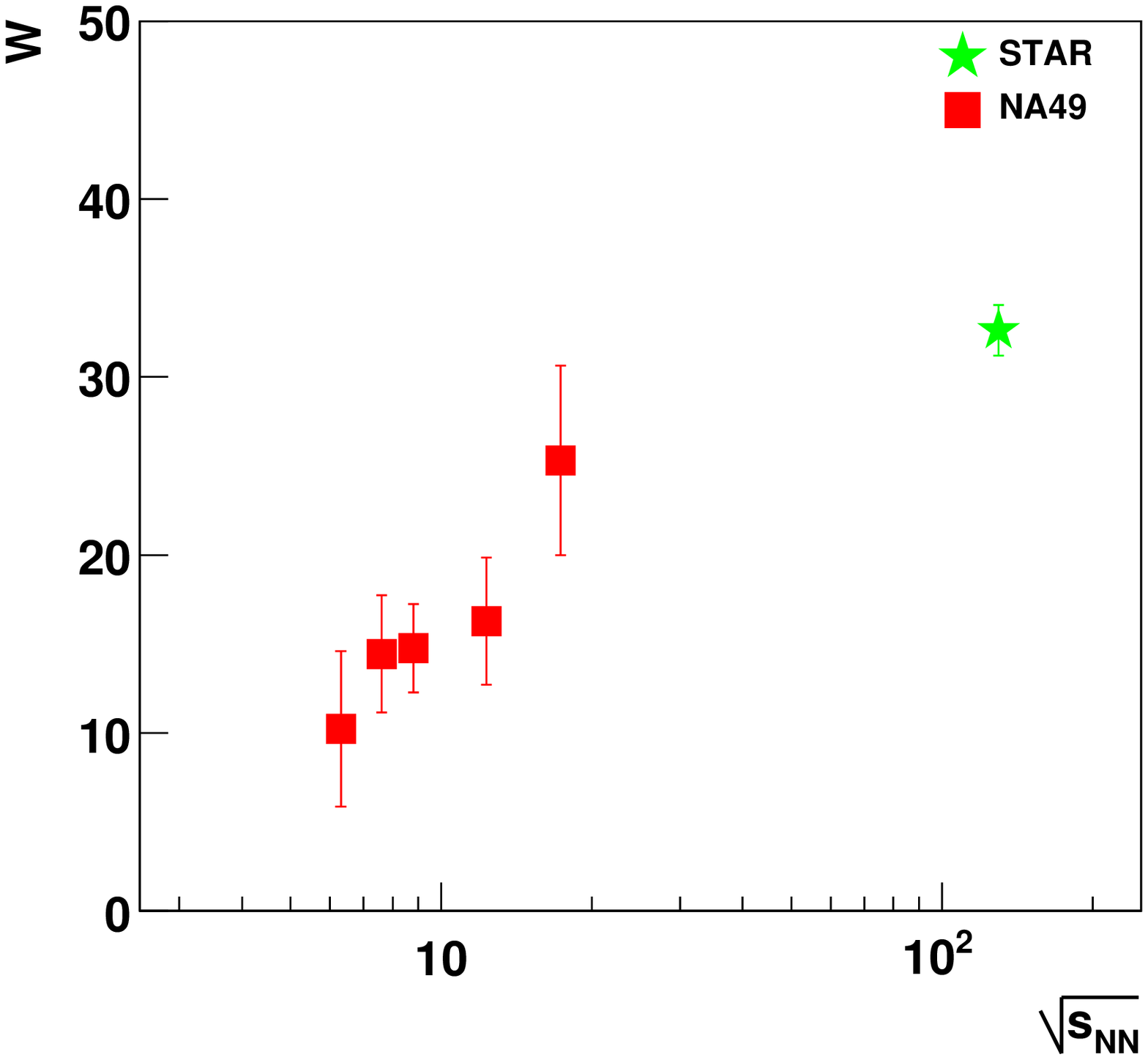}
\caption{The dependence of the normalized width decrease parameter W on $\sqrt{s_{NN}}$ for central Pb+Pb collisions in the SPS energy range after applying the NA49 acceptance filter. SPS energy range (left plot) and extended energy range (right plot) also showing the results from RHIC.}
\label{energy_NA49}
\end{figure}

The left plot of fig. \ref{energy_NA49} shows the dependence of the W parameter on $\sqrt{s_{NN}}$. The data indicate an energy dependence at the level of $(40.4 \pm 19.3)\%$ for our phase-space detector acceptance.

In addition, two models were used in order to further investigate this energy dependence. The Ultra-relativistic Quantum Molecular Dynamics model (UrQMD) \cite{Urqmd} and the Hadron-String Dynamics (HSD) transport approach \cite{Hsd} are two  microscopic models used to simulate (ultra)relativistic heavy ion collisions in the energy range from Bevalac and SIS up to AGS, SPS and RHIC. The points from the corresponding analysis of both UrQMD and HSD generated events throughout the whole SPS energy range can also be seen in fig. \ref{energy_NA49}, where one notices no sign of energy dependence of the W parameter.

The right plot of fig. \ref{energy_NA49} shows the dependence of the normalized width decrease parameter W on the $\sqrt{s_{NN}}$ for the whole SPS energy range as well as for Au+Au collisions at $\sqrt{s_{NN}} = 130$ GeV. The corresponding RHIC point tends to be even higher than the last SPS one, indicating an additional rise of the W parameter towards RHIC leaving an open question for the LHC energies.

\section{Summary}

In this paper, we presented the latest results on the BF obtained by the NA49 collaboration. We have analyzed data coming from different systems and centrality classes for two different energies in central and forward rapidity regions. We have observed a system size and centrality dependence of the width of the BF in the mid-rapidity region which is not seen in the forward region.

Furthermore, the energy scan revealed that data show an indication of an energy dependence which is not reproduced by the HSD and UrQMD models. The normalized width decrease parameter W rises from the lowest SPS energy to the highest SPS and RHIC energies.

\begin{theacknowledgments}

This work was supported by the University of Athens/Special account for research grants, the US Department of EnergyGrant DE-FG03-97ER41020/A000, the Bundesministerium fur Bildung und Forschung, Germany, the Polish State Committee for Scientific Research (2 P03B 130 23, SPB/CERN/P-03/Dz 446/2002-2004, 2 P03B 04123), the Hungarian Scientific Research Foundation (T032648, T032293, T043514), the Hungarian National Science Foundation, OTKA, (F034707), the Polish-German Foundation, and the Korea Research Foundation Grant (KRF-2003-070-C00015).

\end{theacknowledgments}

\bibliographystyle{aipproc}

\end{document}